\documentclass{article}
\usepackage{ijcai18}

\usepackage{times}
\usepackage{xcolor}
\usepackage{soul}
\usepackage[utf8]{inputenc}
\usepackage[small]{caption}

\usepackage{booktabs} 
\usepackage{graphicx}
\usepackage{algorithm}
\usepackage{algpseudocode}
\usepackage{mathtools}
\usepackage{tikz}
\usepackage{pgfplots}
\usepackage{amsmath}
\usepackage{caption}
\usepackage{subcaption}
\usepackage{calrsfs}
\DeclareMathAlphabet{\pazocal}{OMS}{zplm}{m}{n}
\newcommand{\Ta}{\pazocal{T}}

\newcommand\norm[1]{\left\lVert#1\right\rVert}

\title{Semi-Supervised Tensor Factorization \\for Node Classification in Complex Social Networks}

\author{Georgios Katsimpras \and  Georgios Paliouras \\ 
NCSR Demokritos, Ag. Paraskevi, Athens, Greece  \\
gkatsibras@iit.demokritos.gr, paliourg@iit.demokritos.gr}

\begin{document}

\maketitle

\begin{abstract}
This paper proposes a method to guide tensor factorization, using class labels.
 Furthermore, it shows the advantages of
using the proposed method in identifying nodes that play a special role in multi-relational networks, e.g. spammers. Most complex systems involve multiple types of relationships and interactions among entities. Combining information from different relationships may be crucial for various prediction tasks. Instead of creating distinct prediction models for each type of relationship, in
this paper we present a tensor factorization approach based
on RESCAL, which collectively exploits all existing relations.
We extend RESCAL to produce a semi-supervised factorization
method that combines a classification error term with the standard factor optimization process.
The coupled optimization
approach, models the tensorial data assimilating observed
information from all the relations, while also taking into
account classification performance. Our evaluation on real-world social network data shows that incorporating supervision, when available,
leads to models that are more accurate. 


\end{abstract}

\section{Introduction}

In recent years, we are experiencing an enormous growth of online social platforms where complex multi-relational networks take form. Interactions in a multi-relational network provide information that helps to identify the social position of participating parties \cite{ferrante2012sociology}. Hence, identifying individuals of similar social position or role can be modeled as a task of identifying network nodes that have similar relationships \cite{tian2012structural}. In many cases there is also additional information about individuals apart from relationships and interactions.  

Given the richness of information in multi-relational networks, new methods to analyze this information and produce prediction models are needed. In this direction, several approaches have been proposed to deal with the identification of key nodes in multi-relational networks, such as multilinear PageRank \cite{gleich2015multilinear} and the multi-relational version of  hubs and authorities \cite{li2012har}. However, methods that integrate interactions and properties of the nodes, utilizing tensor representations and tensor decomposition have not been studied extensively. 

\begin{figure}[!t]
\centering
\includegraphics[width=3.7in]{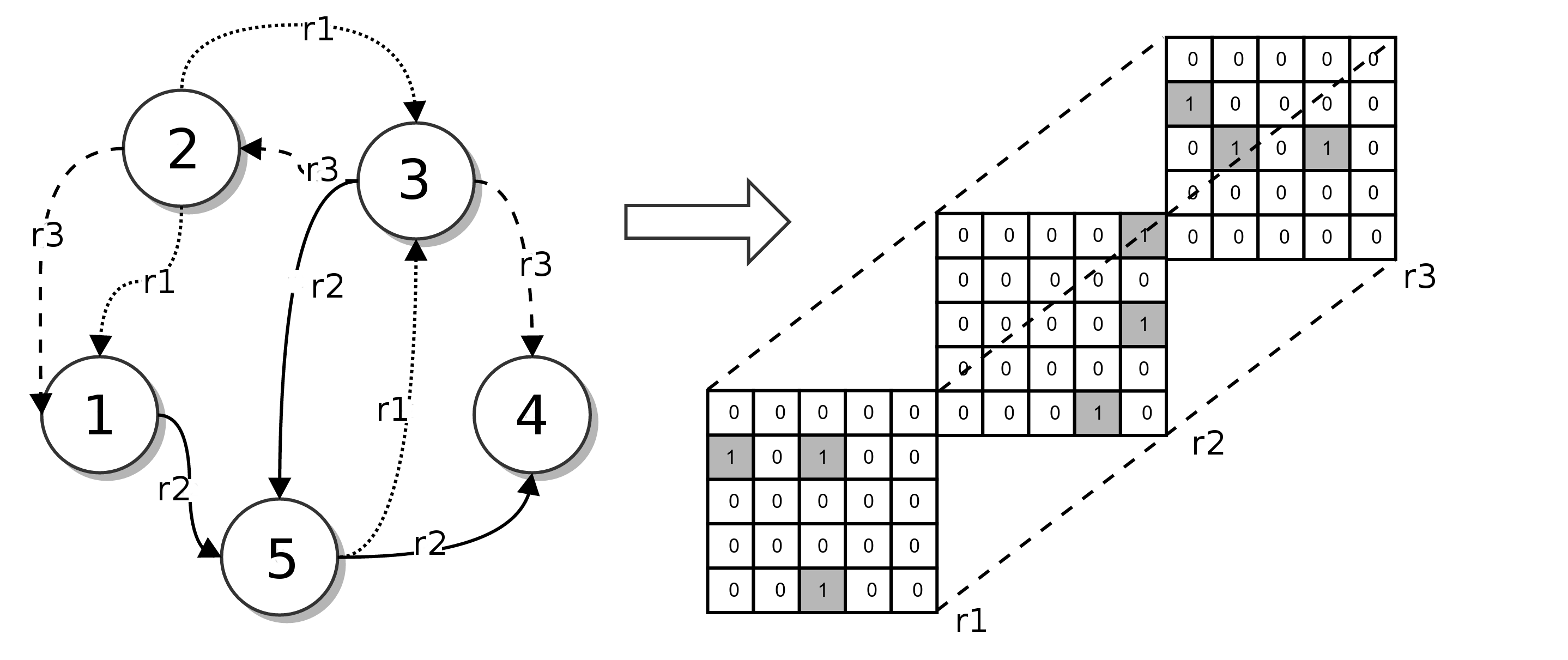}
\caption{A multi-relational network can be easily mapped to a tensor. As we can see, the network at the left can be mapped to a 5x5x3 tensor. Each frontal slice represents the adjacency matrix for a specific relation}
\label{fig_mr_tensor}
\end{figure}

Tensors and their factorization have been used widely in machine learning and data mining. Due to the nature of tensors, relations expressed as tuples of the form \emph{(object, relation, subject)}, can be straightforwardly mapped to a tensor. For instance, the multi-relational network in Figure \ref{fig_mr_tensor}  can be represented as a 5x5x3 tensor where an entry (i,j,r) is 1 only if the relation exists. Rows and columns represent the nodes and each slice/matrix represents the adjacency matrix for a distinct relation.

In this work, we approach the problem of identifying the position and role of individuals in multi-relational networks by representing the networks as tensors and performing tensor factorization. Tensors can provide a simple representation of multi-relational networks while tensor factorization approaches perform well in high dimensional and sparse data \cite{Kolda2009}. Specifically,  we propose a new method for ranking nodes according to their social position in multi-relational networks, combining tensor factorization and classification in a joint learning process. Similar to TripleRank \cite{franz2009triplerank}, our idea is to produce a ranking of users based on the factor scores that are computed during the factorization process. However, in our approach the factor matrices are a result of the proposed semi-supervised factorization process, which forces nodes that belong to the same class to obtain similar factor representation. Thus, ranking in these biased factor matrices will take their class into account.

In order to achieve this, we propose an enhanced version of the RESCAL tensor factorization method that assimilates a-priori information about nodes. Additionally, the factorization process incorporates a component that minimizes classification error. In contrast to existing research that treat factorization and classification separately, we combine them in a single optimization process to obtain semi-supervised factorization. We evaluate the proposed method on a real-wold dataset collected from the social network Tagged.com, that consists of a network of 5,317,649 users and seven types of relations \cite{fakhraei2015collective}.

In summary, the main contributions of the paper are:
\begin{itemize}
\item Combination of tensor factorization with classification,
achieving class-driven modeling of tensorial data.
\item Demonstration of the benefits of using the proposed
method to rank social network users according to their social position, using a real-world dataset
\end{itemize}

\section{Background}
\label{sec_rw}

\subsection{Notation}
Throughout the paper we will use the following notation. The uppercase calligraphic letters denote a tensor  $\Ta \in R^{N\times N\times M}$, as we assume the data is given in such dimensions. Here N is the number of entities (e.g. individuals) and M is the number of relations. Matrices are represented by uppercase italic letters like \emph{A}. Lowercase bold letters like $\mathbf{v}$ denote a vector. The \emph{(i,j)} element of a matrix  \emph{A} is denoted by $a_{ij}$. To refer to the \emph{i-th} row of a matrix  \emph{A} we use $\mathbf{a_i}$. Similarly, an element \emph{(i,j,k)} of a tensor  $\Ta$ will be denoted as $\Ta_{ijk}$. Also, $\Ta_k$ represents the $k$ frontal slice of tensor $\Ta$. Additionally, $vec(\emph{A})$ is the vectorization of  \emph{A} and the operator $\otimes$ is the Kronecker product.

For our relational modeling of triples of the form \emph{(object, relation, subject)}, we use a binary representation: \[ \Ta_{ijk} =
  \begin{cases}
    1       & \quad \text{if the relation exists }\\
    0  & \quad \text{otherwise}\\
  \end{cases}
\] This representation is also known as an adjacency tensor. As an example, the relation \emph{(u1, r2, u5)} in Figure \ref{fig_mr_tensor} exists, causes the corresponding entry in the tensor to be 1, i.e.  $\Ta_{1,5,2} = 1$. In this sense, each frontal slice of $\Ta$ represents an adjacency matrix for all entities with respect to a particular relation.

\subsection{Related Work}
As multi-relational data can be efficiently represented by tensors, many approaches  have been proposed in the literature that employ tensor factorizations. Common tensor factorization approaches such as  CANDECOMP/PARAFAC \cite{harshman1994parafac}, Tucker factorization \cite{tucker1966some} and DEDICOM \cite{bader2007temporal} have been utilized. Most of these studies use a common latent representation for both entities and relations. CANDECOMP/PARAFAC decomposes a tensor into a sum of rank-one tensors. Tucker  decomposes a tensor into a core tensor and separate factor matrices for each mode of the tensor. 

 A recent approach to tensor factorization is RESCAL \cite{nickel2011three} which achieves high predictive performance in the task of link prediction. RESCAL, which we will describe in more detail in Section \ref{sec_rescal}, uses a unique latent representation for entities. 

Recently, several studies have combined matrix and tensor factorization in a joint model. These techniques achieve good performance when there is additional features about the nodes or when combining different datasets together.  A CANDECOMP/PARAFAC based method  for  jointly  factorizing  a  matrix  and  a tensor was proposed in \cite{acar2013understanding}.The matrix was used to hold additional node properties. Similar approaches appear in \cite{narita2011tensor,erdos2013discovering,nimishakavi2016relation}.

\textbf{Identifying Key Nodes in Multi-relational Data.} The problem of identifying key nodes in a network has been extensively studied in the literature. While most of the existing methods apply to single relation networks, there is also great interest in multi-relational networks. The study in  \cite{gleich2015multilinear} is an extension of PageRank to multi-relational networks. In \cite{Liu2010Lu} they propose a graphical model which utilizes heterogeneous link information and the textual information associated to each node. In \cite{zhaoyun2013mining} a method for performing combined random walks  exploiting multi-relational influence networks is presented. MultiRank \cite{ng2011multirank} is a method to simultaneously  determine  the  importance  of  both  objects  and  relations  based  on  a  probability  distribution   computed   from   multi-relational   data. HAR \cite{li2012har} calculates hub,   authority   and   relevance scores of nodes in multi-relational  networks. 
In \cite{jingjing2013mining} a three-stage process is proposed which a) collects features of the users, b) performs a tensor factorization on user relations and c) applies a ranking scheme to identify the most influential users.

As multi-relational data can be efficiently represented by tensors, TripleRank \cite{franz2009triplerank} employs tensor factorization in order to rank entities in the context of linked data. TripleRank applies a common tensor factorization CANDECOMP/PARAFAC \cite{harshman1994parafac} to obtain two factor matrices which correspond to hub and authority scores.  Another approach \cite{rendle2009learning}, utilizes tensor factorization for ranking tags in order to provide tag recommendations, using multi-target networks which involves more than one entity types.

In contrast to these approaches, our method incorporates classification error in the optimization process. In this direction, there are only few approaches that obtain the factor matrices in a supervised manner,that is, by utilizing the label information of the samples. Most of these approaches build a classifier using the factor matrices that are produced by the decomposition \cite{Li2010NonnegativeMA}. Differently from these methods, our approach performs decomposition by estimating the classification error with respect to the factor matrices in each update. Similar to our approach, the method in \cite{wu2013supervised} pursues discriminative tensor decomposition by coupling non-negative Tucker tensor factorization and a maximum margin classifier. In contrast to this method, our approach (a) utilizes RESCAL which does not require the construction of core tensor and (b) produces factor matrices by employing Alternating Least Squares (ALS) which has been proved to be more efficient than gradient descent.


\section{Our Approach: Semi-Supervised RESCAL (CLASS-RESCAL)}
\label{sec_rescal}
\textbf{RESCAL tensor factorization.}
RESCAL is a state-of-the-art relational learning method based on a variant of  DEDICOM 
tensor factorization. It achieves high predictive performance in various tasks such as link prediction and collective classification \cite{nickel2011three}. A basic aspect of RESCAL's model is that an entity has a unique representation over all relations in the data.

Standard tensor factorization models such as CANDECOMP/PARAFAC and Tucker use a bipartite model of relational data, meaning that entities have different latent representations as subjects or objects in relations. Consequently, all direct and indirect relations have a determining influence on the calculation of the latent factors.

RESCAL factorization is computed for each frontal slice as follows:\[\Ta_k = AR_kA^T \text{, for } k = \text{1,..,M} \] where \emph{A} is a $n$x$r$ latent factor matrix that models entities and \emph{$R_k$} is an asymmetric $r$x$r$ latent factor matrix that models relations. More precisely, each entity is represented via a unique row in \emph{A} and each relation via \emph{$R_k$}.

To compute this factorization,  a regularized least squares optimization problem must be solved:
\begin{equation}
\begin{aligned}
& \underset{A, R}{\text{minimize}}
&  \sum_{k} \norm{\Ta_k - AR_kA^T}^2 \\
& & +\lambda_A \norm{A}^2+\lambda_R \sum_{k} \norm{R_k}^2\\
\end{aligned}
\end{equation}
where $\lambda_A$ and $\lambda_R$ are regularization hyperparameters.
This function can be minimized via Alternating Least Squares (ALS) \cite{bader2007temporal}, where the updates of \emph{A} and  \emph{$R_k$} can be calculated as:
\begin{equation}
\begin{aligned}
&A \gets [\sum_{k=1}^M \Ta_kAR_k^T + \Ta_k^T AR_k]\\
&[\sum_{k=1}^M R_kA^TAR_k^T + R_k^TA^TAR_k+ \lambda I]^{-1}
\end{aligned}
\end{equation}
and
\begin{equation}
\label{eq_r}
\begin{aligned}
&R_k \gets (Z^TZ + \lambda I)^{-1} Z vec(\Ta_k)
\end{aligned}
\end{equation}
where $Z = A\otimes A$.
Here, it is important to recall that during the optimization process, a unique latent representation for each entity is shared over all  relations. This means that the resulting \emph{A} takes into account all relations.

\textbf{Introducing classification error.}
We now describe our joint model for tensor factorization and classification. Suppose we have a set of labels for our existing tensorial data. For example, we have a label for some nodes in Figure \ref{fig_mr_tensor} which indicates whether the user is a spammer or not. We could assume that spammers interact and relate with each other \cite{mcpherson2001birds}. In other words, we expect similarly labeled entities to share similar factors.  Based on this assumption, we propose CLASS-RESCAL, to introduce class-label information in the tensor factorization, in order to move entities of the same class closer in the latent space. CLASS-RESCAL models this problem as a joint optimization of tensor factors and classification.

Given an adjacency tensor $\Ta \in \rm I\!R^{N\times N\times M}$, as presented in Section \ref{sec_rw}, a set of labels $Y \in {\{-1,1\}} \text{where } i=\text{1,..,K with K$<$N}$, we solve the optimization problem presented in Eq. \ref{eq_opt}.
\begin{equation}
\label{eq_opt}
\begin{aligned}
& \underset{A, R}{\text{minimize}}
&  f(A,R) + g(A) +h(A,R)\\
\end{aligned}
\end{equation}
where 
 \[ f(A,R) = \sum_{k} \norm{\Ta_k - AR_kA^T}^2\] 
is the tensor factorization least squares problem,
\begin{equation}
\label{eq_knn}
\begin{aligned}
& g(A) = \lambda_g \norm{Y - y(A)}^2
\end{aligned}
\end{equation}
 is the prediction error of the classifier and
  \[h(A,R) =\lambda_A \norm{A}^2+\lambda_R \sum_{k} \norm{R_k}^2 \]
is the regularization term. $\lambda_g$ is a hyperparameter to control the influence of the classification error in the optimization. Note that $y(A)$ is produced with respect to the current values of  the latent factor matrix \emph{A} in each iteration.

\textbf{The Classifier.}
Each row $\mathbf{a_i}$ of the factor matrix \emph{A} represents a unique latent-component representation of the \emph{i-}th entity. The similarity of entities is assumed to be reflected in their latent-component representation. Namely, entities that share similar relations will also share similar latent representations. 
So, if  the  \emph{i-}th entity is similar to  \emph{j-}th entity then also their rows on the factor matrix \emph{A} should be similar, e.g. $\mathbf{a_i} \approx \mathbf{a_j}$. In order to force similar entities of the same class to have similar latent representation we minimize the prediction error of the classifier, Eq. \ref{eq_knn}. 

\begin{figure}[!t]
\centering
\includegraphics[width=2.5in]{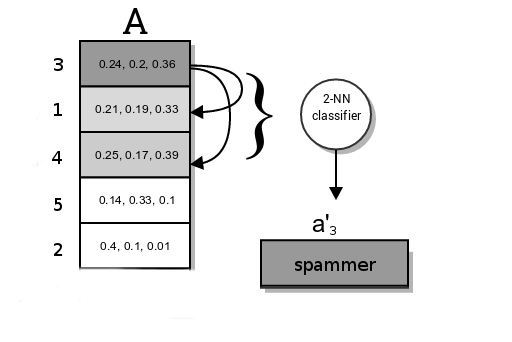}
\caption{The process of k-nn classification. The top-k similar entities are considered for labeling. The labeling is based on majority voting. Here, top-2 nn of node 3 are nodes 1 and 4, who are labeled as spammers (shaded). As a result, node 3 will also be classified as spammer.}
\label{fig_nnc}
\end{figure}

To induce the similarities between entities into the decomposition we use a \emph{k-NN} classifier. Nearest neighbor techniques use the similarity of the local neighborhood of a node to make a prediction, assuming that the social position of a node is close to the social positions of its neighbors. Under this assumption, we assign a label to each row of \emph{A} based on its top $k$ neighbors in \emph{A} (see Figure \ref{fig_nnc}).

For example, assume that Figure \ref{fig_nnc} shows the factor scores of \emph{A} for the network in Figure \ref{fig_mr_tensor}. The latent representation of node 3 is similar to the latent representations of nodes 1 and 4. Assuming that both 1 and 4 are spammers according to supervision, the majority rule will identify user 3 as spammer.

\textbf{Updates.}
\label{sec_update}
We solve the minimization problem in Eq. \ref{eq_opt} by using the efficient alternating least squares method (ALS), as in \cite{nickel2011three}. This approach fixes and solves A and R by update rules, that set the gradient of Eq. \ref{eq_opt} with respect to each matrix to zero.
The update rule for $R_k$ will remain the same compared to the original RESCAL update rule, as in Eq. \ref{eq_r}, while the update rule for A becomes:

\begin{equation}
\label{eq_a}
\begin{aligned}
&A \gets  [\sum_{k=1}^m \Ta_kAR_k^T + \Ta_k^T AR_k + 2(Y - y(A))\\
& [\sum_{k=1}^m R_kA^TAR_k^T + R_k^TA^TAR_k+ \lambda I]^{-1}
\end{aligned}
\end{equation}

When $\frac{f(A,R) + g(A) +h(A,R)}{\norm{\Ta}^2}$ converges to a predefined threshold $\epsilon$  or a maximum number of iterations is exceeded, the procedure stops. The details of the proposed method are summarized in Algorithm \ref{alg_re}.

\begin{algorithm}
\caption{CLASS-RESCAL: Given a tensor $\Ta$ and a set of labels Y, approximate A and R}
\label{alg_re}
\begin{algorithmic}[1]
\Require{adjacency tensor $\Ta$, labels Y, }
\Ensure{latent matrix A, latent matrices $R_k$}
\State Initialize A, R and hyperparameters $\lambda_g, \lambda_A$
\Repeat 
\Procedure {update}{A}
\State $y(A) \gets$ (\Call {Classifier}{A})
\State update A using Eq. \ref{eq_a}
\EndProcedure
\Procedure {update}{R}
\For {each $k$ in $k$-relations}
\State update $R_k$ using Eq. \ref{eq_r}
\EndFor
\EndProcedure
\Until convergence

\end{algorithmic}
\end{algorithm}

\textbf{Identifying entities.}
\label{sec_rank}
In RESCAL, the factor matrix A is a latent representation of users. Each row $\mathbf{a_i}$ contains $r$ scores, one for each latent factor. A ranking of users based on the factor scores will thus prioritize the users on how they are placed on the factor. However, in CLASS-RESCAL the factor matrix A is also influenced by the classifier which forces users of the same class to obtain similar factor representation. Therefore, to identify key users we rank the rows of the factor matrix A. For each row $\mathbf{a_i}$ we compute a ranking score based on the factor scores as follows:
\[ \mathbf{a_i^{score}} = \frac{\sum_{j=1}^r a_{ij}}{r} \]

\textbf{Folding-in a new instance.}
\begin{figure*}[!t]
\centering
\includegraphics[width=5in]{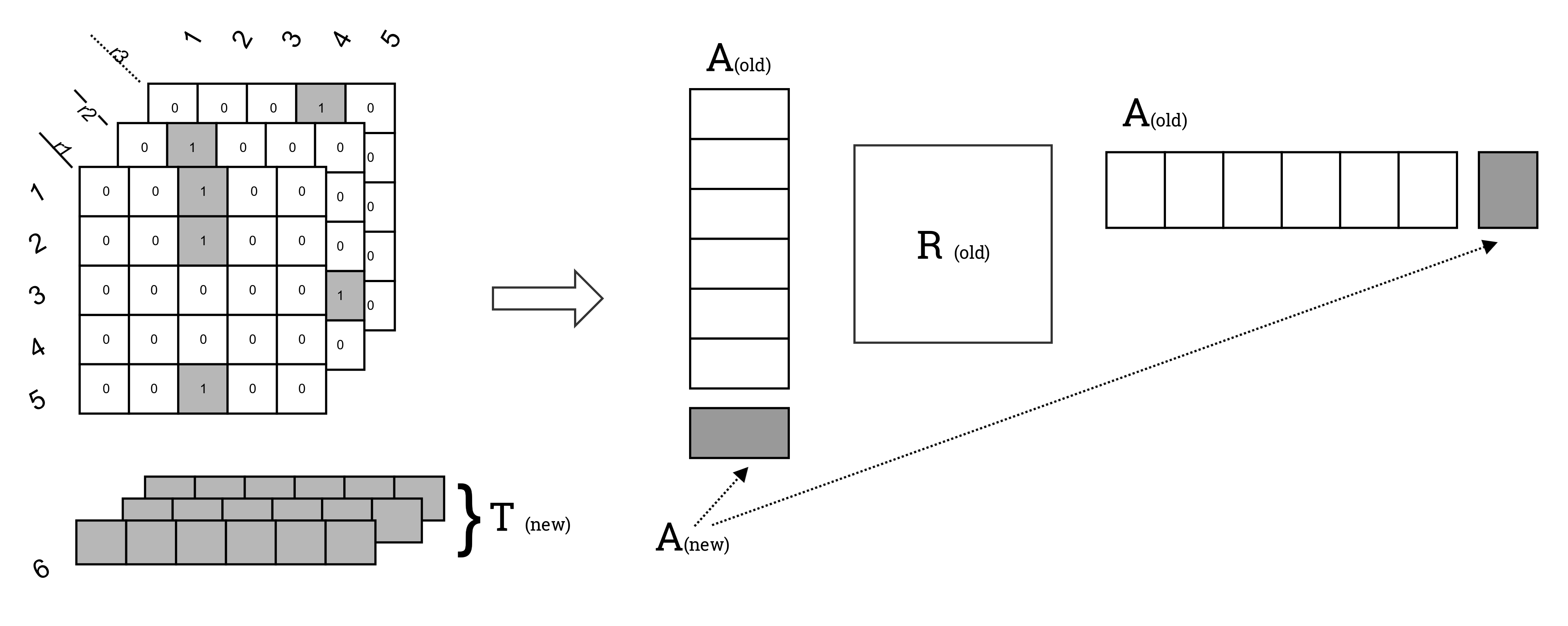}
\caption{The \emph{folding-in} process of CLASS-RESCAL. Note that R remains unchanged}
\label{fig_fold}
\end{figure*}
\label{sec_fold}

Given a new instance, we want to project it to the existing latent space of \emph{A} and \emph{R}. Figure \ref{fig_fold}, illustrates the folding-in process for CLASS-RESCAL. The shaded part of $\Ta$ is the new instance which represents a new user, $\Ta_{(new)}$. \emph{R} remains unchanged, while \emph{A} will have to be updated by the following equation:
\begin{equation}
\begin{aligned}
&A_{(new)} = [\sum_{k=1}^m \Ta_{k(new)}A_{(old)}R_{k(old)}^T + \Ta_{k(new)}^T A_{(old)}R_{k(old)}]\\
&\qquad\qquad [\sum_{k=1}^m R_{k(old)}A_{(old)}^TA_{(old)}R_{k(old)}^T \\
&\qquad\qquad + R_{k(old)}^TA_{(old)}^TA_{(old)}R_{k(old)}+ \lambda I]^{-1}
\end{aligned}
\end{equation}

This procedure is fast requiring just a few simple matrix operations and hence it can be used for the classification of unseen entities. 

\textbf{Complexity.}
Following the analysis in \cite{nickelyago}, the complexity of updating \emph{A} and \emph{R} in RESCAL is $O(pnr + nr^2)$ and $O(pnr^2 + nr^2)$ respectively, where \emph{p} is the non-zero entries, \emph{n} is the number of entities and \emph{r} is the number of factors. Adding a k-nn classifier \footnote{with a kd-tree implementation} in the update process of \emph{A}, will only change the complexity of updating A to $O(pnr + nr^2 + rlog(n))$. 


\section{Experiments}
\label{sec_exp}
\textbf{Dataset.}
We conduct experiments on a real-world dataset. \cite{fakhraei2015collective} was collected from Tagged.com and deals wtih spammer detection. It consists of 5,617,345 users, who interact with each other in seven (7) different ways. The dataset presents high sparsity with the average density being lower than 0.002\%. Due to memory issues, we perform random sampling in order to decrease the size of the dataset. We sample \emph{N} users labeled as spammers and \emph{N} users labeled as non-spammers, and for these users we find all interactions among them that are reported in the dataset. If there are no interactions for a user we discard them. Finally, we come up with a MxMx7 tensor, where M is the number of users left after we remove the users without interactions. For the experiments we set M=6.733.

\subsection{Analysis}
As a performance measure we use the precision-recall curves. Precision is computed as the number of influential users identified at that point relative to the number of points examined. Recall is computed as the number of influentials identified at that point relative to the total number of influentials.

\begin{figure}[!t]
\centering
\includegraphics[width=\linewidth]{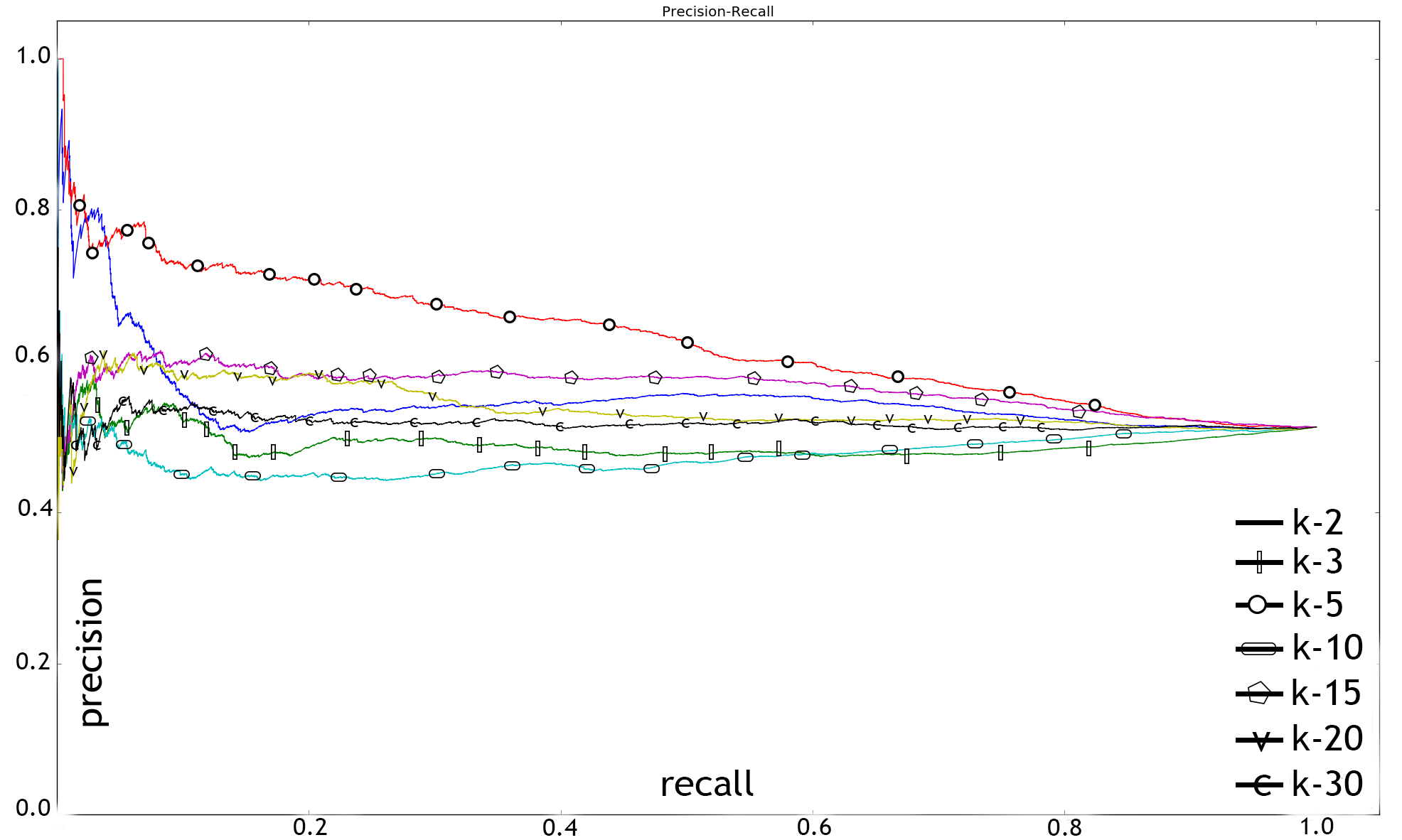}
\caption{Behavior of CLASS-RESCAL while using different number of factors. For our experiments we use r = 5.}
\label{fig_factors}

\end{figure}

\textbf{Number of factors.} Before comparing our method
against other approaches we want to measure the effect of the
number of factors on its performance. We experiment with various r = {2, 3, 5,
10, 15, 20, 30} (see Figure \ref{fig_factors}). A great increase in precision-recall
curve from r = 3 to r = 5. As the number of factors increases, the computation time of the factorizations also increases. As a result for the rest of the experiments we use r = 5 which seems to be a good compromise between performance and computational cost.

\begin{figure}[!t]
\centering
\includegraphics[width=\linewidth]{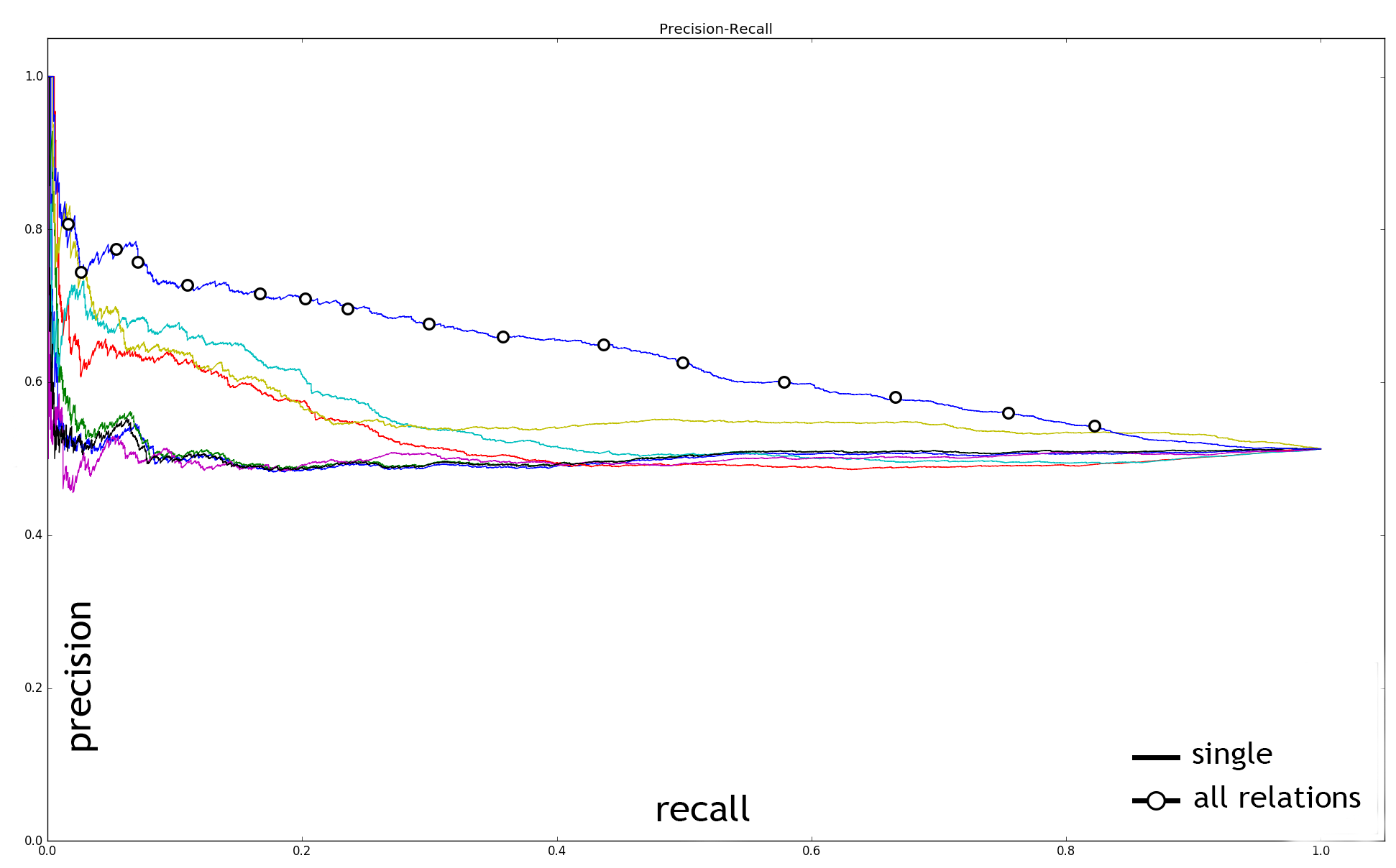}
\caption{Performance of CLASS-RESCAL for various values of the number of relations.}
\label{fig_relations}

\end{figure}

\textbf{Number of relations.}
We also examine the role of the number of relations in the performance of the factorization. In Figure \ref{fig_relations} we show that the performance improves when using all available information (all seven relations).

\subsection{Identifying Key Nodes in Social Networks}
\textbf{Evaluation Methodology.}
In this experiment, we compare the performance of our method against other related approaches in the Tagged.com dataset.
As CLASS-RESCAL is a semi-supervised method, we evaluate it over various training sizes $s \in [10,40]$ percent, averaging the results over 10 random runs. In each run, we sample a random number of instances from each label to use as the training set, and consider the remaining as the test set. With the use of the training set we obtain the semi-supervised tensor decomposition. Then, for each instance in the test set we compute its latent representation using the \emph{folding-in} technique presented in Section \ref{sec_fold}. Thereupon, we compute a ranking with respect to the ranking scores of users, $\mathbf{a_i^{score}}$, as described in Section \ref{sec_rank}. The results reported correspond to the performance of the methods on the test set only.

We employ a widely used metric to evaluate the performance of different methods, namely the area under the precision-recall curve (AUPR).

\textbf{Methods.}
To compare our model, we choose methods that (a) can be applied to multi-relational data and (b) deal with node identification :

\begin{itemize}
\item {\textbf{SVM-degree}: This is a baseline SVM model built by using as features the degree of a node in each relation. }
\item {\textbf{RESCAL}: This is the original RESCAL factorization.}
\item{ \textbf{MultiRank}: MultiRank is a co-ranking method based on random walks in multi-relational data. It constructs two transition probability tensors, one for objects and the other for relations. Based on these tensorial representation of multi-relational data it performs a pagerank like random walk and computes a separate multirank score for objects and another for relations.}
\item{ \textbf{HARrank}: This is a multi-relational analog of the HITS algorithm. It constructs three different transition probability tensors, one for hubs, one for authorities and the last one for relations. Similar to MultiRank, it performs a random walk in the created tensors and computes three different scores (hub, authority, relation).}
\item{ \textbf{TripleRank}:  This is a CANDECOMP/PARAFAC based tensor factorization model for ranking entities in linked data. It uses the factor scores produced from the tensor decomposition to rank entities.}
\end{itemize}

In the experiments, we set the necessary parameters for each method. For our method, we set the number of factors $r=5$ and hyperparameters $\lambda_g=0.1$ and $\lambda_A,\lambda_R=0.5$, after performing grid search. We also use $r=5$ for TripleRank. For the rest methods we use the values proposed by the authors.

\textbf{Experimental Results.}
Table \ref{tab_auc} shows the performance of all methods on the Tagged.com dataset. CLASS-RESCAL outperforms the other methods, while RESCAL yields the lowest performance. SVM-degree performs better than RESCAL, which indicates the importance of supervision. MultiRank seems to perform slightly better than the other methods. In conclusion, the use of class-label information in the decomposition of CLASS-RESCAL, seems to lead to a more effective ranking, confirming our original intuition.

\begin{table}[!t]
\caption{The AUPR curve scores for all methods.}
\label{tab_auc}
\centering
\begin{tabular}{|c||c|c|c|}
\hline
  AUPR &  Tagged.com  \\
\hline
\hline
SVM-degree  &0.51 \\
\hline
RESCAL  & 0.46\\
\hline
MultiRank  & 0.52\\
\hline
HARrank  & 0.49\\
\hline
TripleRank  & 0.51\\
\hline
CLASS-RESCAL &\textbf{ 0.64}\\
\hline

\end{tabular}
\end{table}

\begin{figure}[!t]
\centering
\includegraphics[width=\linewidth]{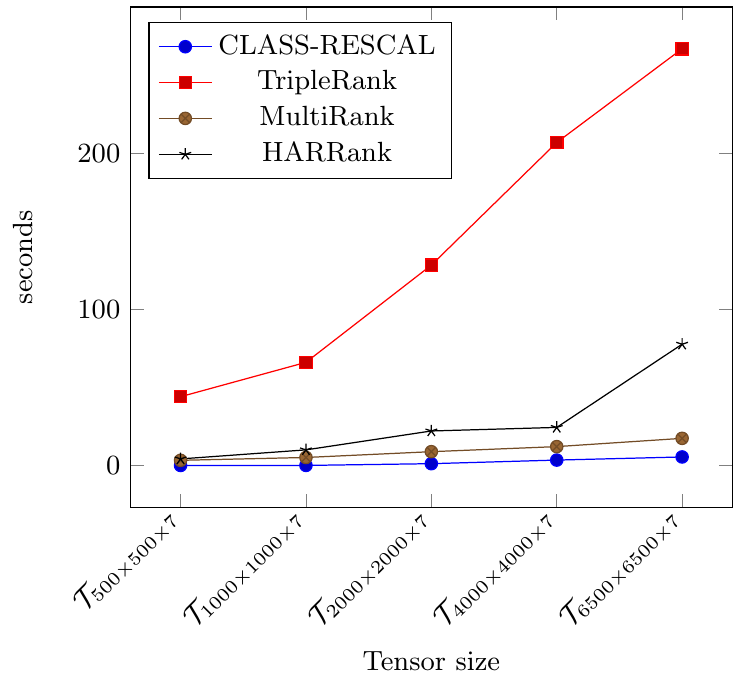}
\caption{Running times of all methods for different tensor sizes.}
\label{fig_rtimes}

\end{figure}

\textbf{Running Times.}
For the runtime comparison we performed various tests experimenting with the size of the tensors.
An empirical runtime comparison for predicting a ranked list of users is depicted in Figure \ref{fig_rtimes}. As we can see, the runtimes of all methods are dependent on the size of the tensor and hence on the size of the dataset. As the size increase, the runtime of all methods increases. However, the runtime of CLASS-RESCAL differentiates from the rest. The runtimes of MultiRank and HARRank are reasonable since they perform random walks on existing edges which results to lower running times when the sparsity of the dataset is high. The runtime of TripleRank is problematic even for small datasets and becomes unfeasible as the size of the tensor increases.   

\section{Conclusion - Discussion}
\label{sec_con}

In this paper, we presented an extension of the tensor factorization method RESCAL, for identifying the social position of users in a multi-relational network by employing supervision. 
The factorization we proposed incorporates a classification error function in the optimization process achieving class-driven modeling
of the tensorial data.  Our method models the tensorial data while minimizing classification error.
Therefore we obtain a semi-supervised tensor factorization which assimilates class-label information. 
In our evaluation, we conducted experiments with ground truth data that demonstrate the usefulness of labeled data for tensor factorization and illustrate the effectiveness of our approach. 

Among our immediate plans to extend this work, we are studying other tasks of Social Network Analysis, like link prediction and recommendation. At the same time we are exploring different tensor factorization and classification methods. Additionally, we plan to evaluate our approach extensively with more datasets that contain more types of relations.

\section*{Acknowledgments}
This paper is supported by the project "HOBBIT- Holistic Benchmarking of Big Linked Data", which has received funding from the European Union’s Horizon 2020 research and innovation programme under grant agreement No 688227.
\bibliographystyle{named}
\bibliography{sigproc}

\end{document}